\documentclass[11pt]{article}
\pdfoutput=1
\usepackage{cite}
\usepackage{amsmath,amssymb,amsfonts}
\usepackage{graphicx}
\usepackage{textcomp}
\usepackage{url}
\usepackage{xcolor}
\usepackage{subcaption}
\usepackage{algorithm,algorithmicx,algpseudocode}
\usepackage{colortbl}
\usepackage{lipsum}
\usepackage{float}
\usepackage{pifont}
\usepackage{tabularx}
\usepackage{mwe}

\begin{document}

\title{Does Link Prediction Help Detect Feature Interactions in Software Product Lines (SPLs)?}

\author{Seyedehzahra Khoshmanesh and Robyn Lutz\\
       Department of Computer Science\\
       Iowa State University\\
       Ames, IA USA 50011\\
      \{zkh,rlutz\}@iastate.edu\\
     }

\maketitle

\thispagestyle{plain}
\pagestyle{plain}

\begin{abstract}

An ongoing challenge for the requirements engineering of software product lines is to predict whether a new combination of features (units of functionality)  
will create an unwanted or even hazardous feature interaction.
We thus seek to improve and automate the prediction of unwanted feature interactions early in development.  In this paper we show how the detection of unwanted feature interactions in a software product line can be effectively represented as a link prediction problem.   Link prediction uses machine learning algorithms and similarity scores among a graph's nodes to identify likely new edges.  We here model the software product line features as nodes and the unwanted interactions among the features as edges.  We investigate six link-based similarity metrics, some using local and some using global knowledge of the graph, for use in this context.  We evaluate our approach on a software product line benchmark in the literature, building six machine-learning models from the graph-based similarity data. Results show that the best ML algorithms achieved accuracy of 0.75 to 1 for classifying feature interactions as unwanted or wanted in this small study, and that global similarity metrics performed better than local similarity metrics.  The work shows how link-prediction models can help find missing edges, which represent unwanted feature interactions that are undocumented or unrecognized, earlier in development. 

\end{abstract}

\section{Introduction}

\begin{figure*}
  \centering
  \includegraphics[width=\linewidth]{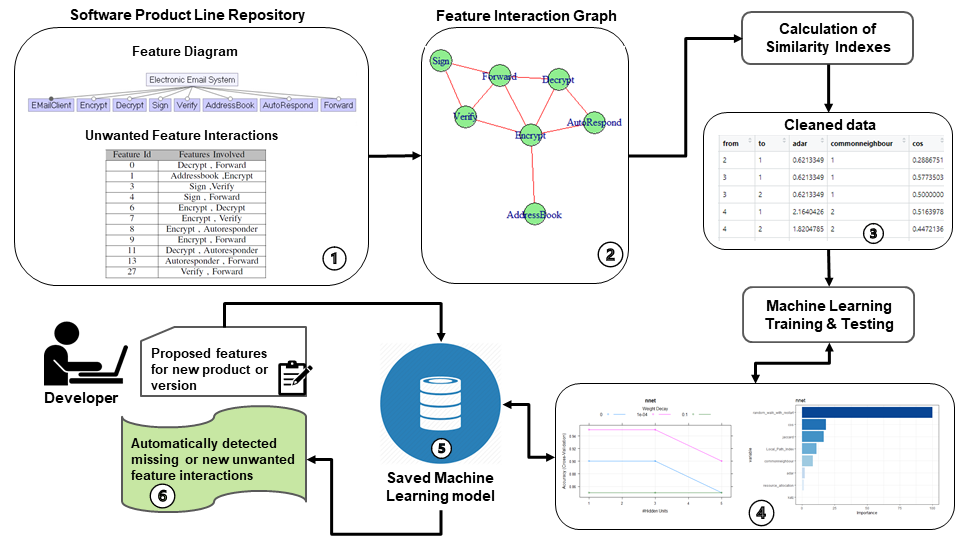}
  \caption{Workflow of proposed method to classify unwanted feature interactions}
  \label{fig:framework}
\end{figure*}

Software product lines are widely used in industry to reap the benefits of reuse.  
A software product line (SPL) is a family of software products that share a set of basic features as a core and differ in other alternative or optional features \cite{pohl2005software}. A {\it feature} in a software product line is a unit of functionality that provides service to users \cite{soares2018feature} (i.e., different from a feature in machine learning or statistics). Features are problem-oriented and describe the users' requirements \cite{classen2008s, bosch2000design}.  

A software product line tends to evolve as it grows.  As more products join the product line over time, new combinations of features get added \cite{botterweck2014evolution}. 
However, some features are incompatible, and can even cause hazardous conditions when combined in a single product \cite{apel2013feature}.  These constraints are termed unwanted feature interactions. 

For instance, an unwanted feature interaction occurs in a telephony system when we combine the two  features call-forwarding and call-waiting \cite{calder2003feature}. If we enable both features, the system enters an unexplored and unsafe state when, while the line is busy, the system receives another call. In this case there is a requirements conflict and the system does not know whether the call should be delayed or forwarded.

An ongoing challenge for the requirements engineering of software product lines is to predict whether a new combination of features 
will create an unwanted or even hazardous feature interaction.  Detecting such unwanted feature interactions is a difficult and persistent problem for software product lines. Often they are not found until testing \cite{henard2014bypassing} or operations.  While model checking approaches can catch some unwanted feature interactions earlier, they have been difficult to implement at industrial scales \cite{siegmund2012predicting}. 

The work reported here explores a new approach to earlier detection of unwanted 
feature interactions, inspired by link prediction in  networks.  Many social, information and biological systems and networks can be represented as graphs.  For example, a social network is a graph in which each edge shows a friendship between two people (i.e., nodes) in the graph, and a co-authorship network is a graph in which each edge shows a paper collaboration between two authors. 
Links in the field of link prediction are between nodes of the same type, e.g., two people, while
links in the field of traceability are typically between different types of software artifacts, e.g., a requirement and its source code \cite{ValeAAKNL17}.)
Link prediction then uses similarity based algorithms to predict the likelihood of the creation of a new edge between two nodes in the graph. That is, link prediction detects  potential missing links between nodes, such as a missing but likely friendship in a social network \cite{liben2007link,nassar2019pairwise}.

In this paper we show how the detection of unwanted feature interactions in a software product line can be effectively represented as a link prediction problem. We thus model a software product line as a graph of features and relationships or interactions existing between the features. 
Each feature in a software product line feature model is represented as a node in the feature interaction graph.  The links or edges between features  represent the {\it feature interactions} between them.

The work reported in this paper employs the knowledge of prior wanted and unwanted feature interactions captured in a product line's feature model and feature constraints, together with similarity measures among product-line features, link prediction, and machine learning algorithms to
improve and automate the detection of missing or new unwanted feature interactions in a new product. As shown in Fig. \ref{fig:framework} and described in Sect. 2, we apply link prediction techniques to calculate local and global similarity among the features in a feature interaction graph. Next, we build, train, and tune a machine learning model to detect potential new or missing unwanted feature interactions in the new product or version While previous approaches have succeeded at detecting unwanted feature interactions during testing, our approach can find many of them earlier, in the requirements phase. 

Similarity is a key metric in our proposed framework and acts as a heuristic tool for detecting a new or missing unwanted feature interaction.  This is because similar features 
have been observed to behave in similar ways. If there is a feature in the feature interaction graph that contributes to some unwanted feature interactions, the features which are similar to this feature often will contribute to the same unwanted interactions \cite{khoshmanesh2018role,khoshmanesh2019feature,khoshmanesh2019leveraging}.

We thus target two goals in our paper. First, we want to understand whether information about the feature interaction graph can suffice to detect potential missing or new unwanted feature interactions. Second, we want to investigate whether link prediction and machine learning algorithms can help achieve this detection. To address these issues, the paper aims to address the following questions:

\begin{itemize}
    \item \textit {\textbf{RQ1:} How effectively does link prediction help detect unwanted feature interactions in a software product line?} 
    
    \item \textit {\textbf{RQ2:} Which similarity metrics and machine learning algorithms perform better in the context of unwanted feature interaction detection?}

\end{itemize}

We investigate these research questions by applying our approach to a case study, the Electronic Mail system  introduced by Hall in \cite{hall2005fundamental} and extended as a benchmark in the software product line literature \cite{apel2013strategies}. 

Results obtained from the application of our approach to
the Email benchmark showed a perfect accuracy of $100\%$ in detecting unwanted feature interactions using link prediction techniques with Random Forest, Naive Bayes and Linear Support Vector Machine. This indicates that the use of link prediction, similarity metrics, and machine learning algorithms in a software product line may help detect missing or new unwanted feature interactions in the requirements phase of a proposed new product in the product line.

The contribution of the paper is a framework which combines link prediction and machine learning techniques to detect  unwanted feature interactions in the early-phase development of a new product in a software product line or of a new version of a software-intensive system. While similarity measures and link prediction have been considered widely in software testing and social network systems, to our knowledge they have not been studied previously for detection of unwanted software feature interactions.  

The work described in this paper is part of our ongoing effort toward improved detection of feature interactions during the requirements analysis of a new product.  In earlier work we studied similarity measures based on features' structural elements (classes, attributes and methods) \cite{khoshmanesh2018role} and on features' relative positions in a software product line's feature model \cite{khoshmanesh2018role,khoshmanesh2019feature,khoshmanesh2019leveraging}. New work that is first reported in this paper is our representation of the feature interaction problem as a link prediction problem, which makes feature interactions  amenable to classification (wanted/unwanted) and the building of a predictive learning model, together with the evaluation results from our initial application of this approach. 

The rest of the paper is structured as follows. Section 2 describes our similarity-based machine-learning method for detecting unwanted feature interactions in a software product line. Section 3 describes results from an evaluation of its application on a small software product line. Section 4 reviews related work, and Section 5 gives concluding remarks. All artifacts, code, and analysis used in this study are available at
\url{https://tinyurl.com/y8h5erwp}.

\section{Method}
In this section, we explain our proposed method for detecting unwanted feature interactions in the requirements phase of a new SPL product or version.   We first give an overview of the method, as well as of the intuition on which it is based. We then introduce the software product line case study that we use to evaluate this approach. Finally, we define the similarity metrics whose calculated values are used by the learning algorithms.  

\subsection{Overview}
We define a Software Product Line as a graph $G=(V,E)$ in which each feature $F$ in the software product line feature model is a node, $V$, in the graph G. The edge, $E$, between two features $F_{i}$ and $F_{j}$ represents a known feature interaction, either wanted or unwanted, as documented in feature model constraints, between two features, 

\[Feature Interaction=(F_{i}, F_{j})\]
\[G_{Software Product Line}=(V_{F},E_{Feature Interaction})\]

Fig. \ref{fig:framework} shows the framework of our proposed method. We briefly describe the steps in its process, identifying each by its number given in the figure. In \textbf{(1)}, we gather the requirement level artifacts, including the software product line feature model and its associated list of existing wanted and unwanted feature interactions from the software product line repository, to pre-process and create a feature interaction graph, shown in \textbf{(2)}. The feature interaction graph is the appropriate input on which to apply the link prediction technique. In a feature interaction graph, each edge can be labeled {\it wanted interaction} or  {\it unwanted feature interaction}. 

The interaction graph shown in Fig. 3 
is of the Electronic Mail software product line, the case study we use to investigate  our research questions \cite{hall2005fundamental,apel2011detection}.  It is introduced below. The graph 

is automatically created by the ``igraph'' package \cite{csardi2006igraph}. 

\begin{figure}[ht]
\centering
\includegraphics[width=\linewidth]{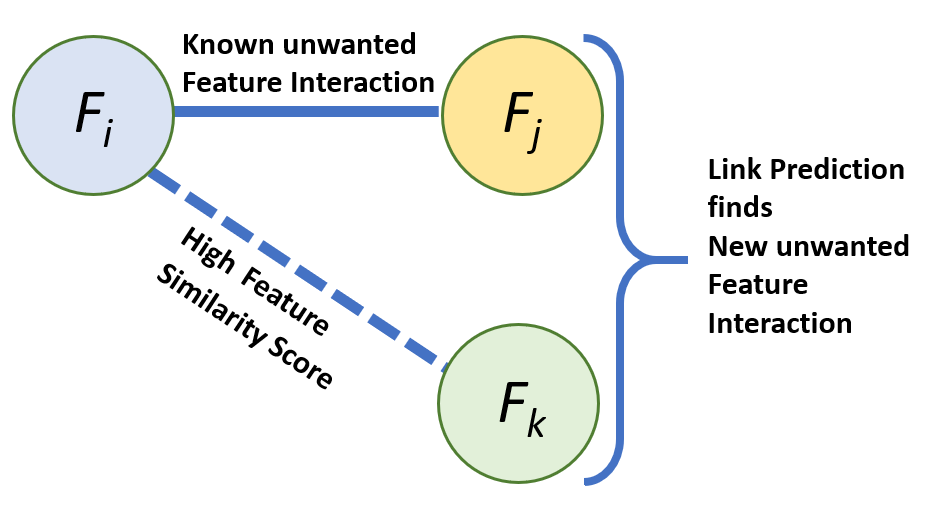}
\caption{Intuition behind our similarity-based learning  to detect unwanted feature interaction}
\label{fig:intuition}
\end{figure}

In the process step \textbf{(3)} of Fig. \ref{fig:framework}, we apply proximity-based methods for link prediction on the feature interaction graph 
to obtain the similarity scores for the feature interaction pairs in the graph.

Fig. \ref{fig:intuition} shows the intuition behind how using similarity indexes between two nodes in an interaction graph helps to detect new or missing unwanted feature interactions in a new version or product-line product \cite{khoshmanesh2018role,khoshmanesh2019feature,khoshmanesh2019leveraging,khoshmanesh2019master}. As shown in Fig. \ref{fig:intuition}, there is an unwanted feature interaction between two features, $F_{i}$ and $F_{j}$ that is shown by the edge between them. If, in a new version or product of a software product line, a new feature $F_{k}$ is added that has a high feature similarity score with feature $F_{i}$, it is similarly likely to have an unwanted interaction with feature $F_{j}$.

To make the intuition described in Fig. \ref{fig:intuition} more concrete, we extend a classical example of unintended feature interaction, described by Batory et al. and attributed to Kang \cite{batory2011feature}.  Suppose our building control product line has three features, each of which operates correctly in isolation.  
$F_{i}$ is a Flood-Control feature with water sensors that, when they detect standing water, turn off the water main. 
$F_{j}$ is a Fire-Control feature with sensors that, when they detect fire, activate water sprinklers. 
There is a known unwanted feature interaction between the Fire-Control feature and the Flood Control feature, shown in Fig. \ref{fig:intuition} as an edge between them.  This is because the features interfere with each other and create a hazardous situation, turning the water main off while the sprinklers should be active.
The third feature, $F_{k}$, is a Pipes-Protection feature with sensors that, when they detect sub-freezing temperatures in the building, turn off the water main to avoid burst pipes. 

Suppose that the Pipes-Protection feature is being added to a building having a Fire-Control feature for the first time. There is no known feature interaction between these two features as they have never been combined in a product previously. 
However, we seek to use the high degree of similarity between the Flood-Control feature and the Pipes-Protection feature, both of which turn off the water main, to  predict that the Pipes-Protection feature may have an unrecognized and unwanted feature interaction with the Fire-Control feature. This previously unrecognized feature interaction can then be suggested to the requirements analyst. If confirmed, it can be documented by adding it as an edge in the graph to also help with future products. 

More specifically, the link prediction techniques uses similarity scores to predict the missing links, i.e., edges, between nodes in a graph \cite{lu2011link}. The link prediction method predicts the new or missing edges, including the edge between $F_{k}$ and $F_{j}$. A similarity score $s_{F_{1} F_{2}}$ in an interaction graph is defined as ``how much'' two nodes in the graph are similar. A higher similarity score means a higher likelihood that the link will appear in the future. We formalize the calculation of similarity below.

\begin{table}[ht]
  \caption{Local and global link-based similarity metrics used in detection of unwanted feature interactions}
  \label{tab:link-based}
  \begin{tabular}{ccc}
    \hline
    & Metric Name & Category\\
    \hline
    1& Common Neighbors & Local Similarity \\
    2& Jaccard distance \cite{jaccard1901etude} & Local Similarity \\\
    3& Cosine distance \cite{singhal2001modern} & Local Similarity \\\
    4& Adar index \cite{adamic2003friends} & Local Similarity \\
    5& Resource Allocation Index (RA) \cite{zhou2009predicting}  & Local Similarity \\
    6& Katz \cite{katz1953new} & Global Similarity \\
    7& Random Walk with Restart (RWR) \cite{sun2005neighborhood}& Global Similarity \\
    8& Local Path Index (LP) \cite{zhou2009predicting} & Quasi-local methods \\
    \hline
\end{tabular}
\label{tab:sim metric}
\end{table}

We use similarity based algorithms to detect missing and new unwanted feature interactions when the software product line evolves, such as features being added or new versions introduced. Table \ref{tab:sim metric} shows the eight similarity metrics which we used in our investigation. 
These similarity metrics can capture local and global similarity between two nodes in the feature interaction graph of a software product line. The similarity metrics that only need the local topology of the graph to be calculated belong to the local similarity category, while the similarity metrics that require global topological information for a graph (e.g., shortest path) belong to the global similarity category. These similarity metrics are  widely used in the link prediction literature, and research studies report that they are among the highest accuracy local and global similarity metrics in fields as varied as network science, electrical power-grids, and protein-protein interaction
networks \cite{rawashdeh2015similarity,zhou2009predicting,lproximity,liu2007state}.

Step \textbf{(4)} in our framework process is to input the data in the form of a data frame to the machine learning models. Each record in our cleaned data describes different similarity scores for an edge between two features in the graph. The class variable indicates whether this edge contributes to an unwanted feature interaction or not.  Therefore, we have a classification or supervised learning problem. We train and tune six different machine learning algorithms, as described below in the Results section. The best machine learning model can then be saved to evaluate on unseen data in the future. We save the optimized final machine learning model in step \textbf{(5)}.

Finally, the requirement analyst can use the saved model shown in \textbf{(5)} of Fig. \ref{fig:framework} to check combinations of product-line features proposed for a new product as early as possible in order to learn about the possible missing or new unwanted feature interactions. The final report, shown in \textbf{(6)}, provides useful information regarding potential missing and new unwanted feature interaction in the new products. Additionally, as a product evolves over its life cycle, the results of applying our method could be used as an incremental learning model in order to improve model accuracy and generalizability.

\subsection{Software product line case study}

In this subsection, we describe the case study on which we applied and evaluated our proposed method for detecting unwanted feature interactions early in the development.   We selected the Electronic Mail System, or Email, software product line from the literature, since it provides multiple feature interactions.  The Email system was originally introduced by Hall \cite{hall2005fundamental} and later became a product-line benchmark used by Apel and others \cite{apel2013feature,apel2013strategies}. The Email software product line models an e-mail communication system having several optional features that can be enabled or disabled such as encryption, forwarding, and verify email.  Its feature model is shown as part of the software product line repository at the top left in Fig. \ref{fig:framework}. It shows eight available features for any new product. The leftmost feature there, ``Email Client,'' is a commonality that must be present in all products.  The other seven features are optional.

\begin{figure}[ht]
\centering
\includegraphics[scale=0.8]{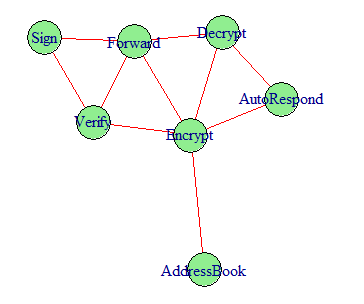}
\caption{Unwanted Feature Interaction Graph for the Email system (automatically created with the ``igraph'' tool in R) }
\end{figure}

Selecting some pairs of optional features will cause unwanted feature interactions. Fig. 3 is an unwanted feature interaction graph for the Email product line. It shows the seven optional features as nodes and the unwanted feature interactions as links, or edges.  There are 10 unwanted feature interactions in the Email product line. 

An example of an  unwanted feature interaction comes from an Email product line \cite{apel2011detection} in which its Encrypt Email feature and its Forward Email feature each work as intended when only one of them is present.  However, when both these features appear in a product, unwanted behavior occurs. Namely, an encrypted email will be forwarded in plain text if the client's public key is not available, violating the product's security requirement.

As we can see in Fig. 3,

``Encrypt'' has the highest degree among features in the interaction graph, as it participates in five unwanted feature interactions (edges) in the graph.  Feature ``Forward'' similarly contributes to four interactions.  

Note that the figure does not display the 11 normal, i.e.,  wanted, feature interactions, that would appear in a fully connected feature interaction graph.  However, we will use all 21 edges (10 unwanted and 11 wanted interactions) of the fully connected graph to build our models. 

\subsection{Similarity Measures to Detect Feature Interaction}
We next describe the similarity metrics used in our investigation.  All the similarity metrics described here have shown good performance even on very large data and have polynomial time complexity \cite{rawashdeh2015similarity}. We selected both local and global similarity indexes. The local similarity indexes are node-based topological similarity metrics, while global similarity indexes are path-based topological similarity metrics.

\subsubsection{Local Similarity Metrics}
\begin{itemize}

       \item Common Neighbors. Two nodes x and y in a graph are more likely to have a link if they have many common neighbours. In the context of feature interaction, two features $F_1$ and $F_2$ are more likely to have an unwanted feature interaction if these two features have many common neighbours in the feature interaction graph.
       For a feature $F_1,$ let $\Gamma(F_1)$ denote the set of neighbors of $F_1 .$ The CN (Common Neighbor) is defined as counting number of in-common neighbors of two features in the graph.
       $$ s_{F_1 F_2}^{C N}=|\Gamma(F_1) \cap \Gamma(F_2)|$$ 
       Many studies have used CN and observed that there is a positive correlation between the number of common neighbours in a graph and possible links between two nodes in a graph, such as in a friendship graph or a scientific collaborative graph \cite{kossinets2006effects,newman2006structure}.

       \item Jaccard Index \cite{jaccard1901etude}. The Jaccard Index measures the similarity between two sets, and is defined as:
       
       $$s_{F_1 F_2}^{J a c c a r d}=\frac{|\Gamma(F_1) \cap \Gamma(F_2)|}{|\Gamma(F_1) \cup \Gamma(F_2)|}$$

        \item Cosine Distance
        \cite{salton2003information}. Cosine distance is a metric used to measure how similar nodes are irrespective of their degree. Mathematically, it measures the cosine of the angle between two vectors projected in a multi-dimensional space. It is defined as:
        $$s_{F_1 F_2}^{\text {Cosine}}=\frac{|\Gamma(F_1) \cap \Gamma(F_2)|}{\sqrt{k_{F_1} \times k_{F_2}}}$$

       where $k_{F_1}$ is the degree of node $F_1 .$ 
       
       \item Adamic/Adar (AA) Index \cite{adamic2003friends}. An AA measure is defined as the inverted sum of degrees of common neighbors for two given vertices. This metric measures the closeness of two nodes based on their shared neighbors.  A value of 0 indicates that two nodes are not close, while higher values indicate that nodes are close. This index refines the simple counting of common neighbors by assigning the less-connected neighbors more weight, and is defined as
       $$ s_{F_1 F_2}^{A A}=\sum_{z \in \Gamma(F_1) \cap \Gamma(F_2)} \frac{1}{\log k_{z}}$$
       
      \item Resource Allocation Index (RA) \cite{zhou2009predicting}. This index is motivated by the resource allocation in networks. Given a pair of nodes, $x$ and $y$ which are not directly connected, the node $x$ can send some resource to $y$ with their common neighbors as transmitters.  We assume that each transmitter has a unit of resource, and will equally distribute it to all of its neighbors. The similarity between $x$ and
      $y$ is defined as the amount of resource that $y$ received from $x$ which is:
      $$s_{x y}^{R A}=\sum_{z \in \Gamma(x) \cap \Gamma(y)} \frac{1}{k(z)}$$

\end{itemize}

\subsubsection{Global Similarity Metrics}
\begin{itemize}

    \item Katz Index \cite{katz1953new}. The Katz centrality of a node is a measure of centrality in a network. Unlike typical centrality measures which consider only the shortest path between a pair of nodes, Katz centrality measures influence by taking into account the total number of walks between a pair of nodes. It is similar to Google’s PageRank.

   \item Random Walk with Restart (RWR) \cite{lu2011link}. This index is a direct application of the Page Rank algorithm.

   \item Local Path Index (LP) \cite{zhou2009predicting}. This metric uses the local paths and wider common neighbors (neighborhoods of second order) to reduce the complexity of the Katz metric. 

\end{itemize}

\section{Results}
In this section, we describe the evaluation results for the two research questions presented in the introduction regarding our method. To answer these research questions, we applied the eight link-based similarity metrics in Table \ref{tab:link-based} to the feature nodes in the feature interaction graph of the Email case study. These similarity metrics are used widely in link prediction literature, and research studies have shown their usefulness \cite{lu2011link}.

\begin{figure*}[ht]
\centering
\begin{subfigure}[b]{0.45\textwidth}
\includegraphics[width=\textwidth]{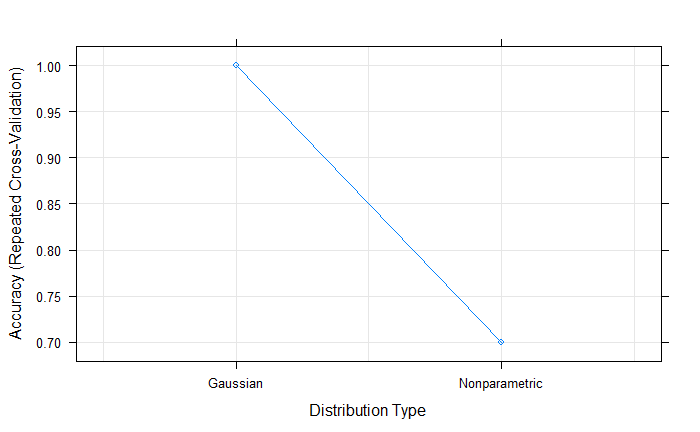}
\caption{Naive Bayes}
\end{subfigure}
\begin{subfigure}[b]{0.45\textwidth}
\includegraphics[width=\textwidth]{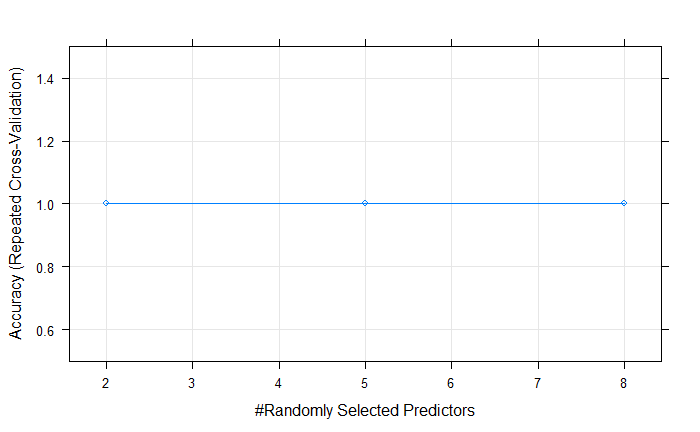}
\caption{Random Forest}
\end{subfigure}
\begin{subfigure}[b]{0.45\textwidth}
\includegraphics[width=\textwidth]{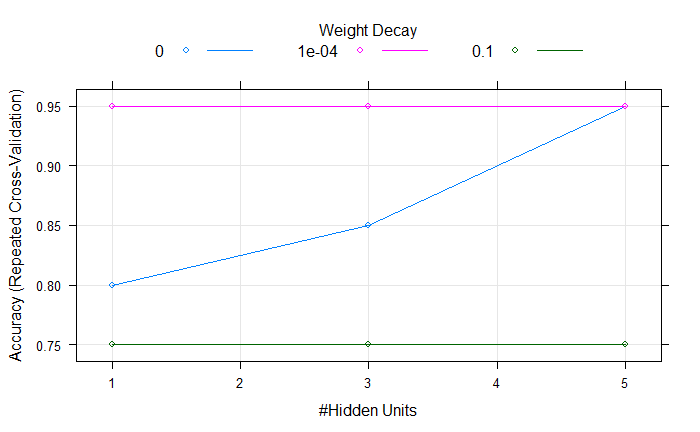}
\caption{Neural Net}
\end{subfigure}
\begin{subfigure}[b]{0.45\textwidth}
\includegraphics[width=\textwidth]{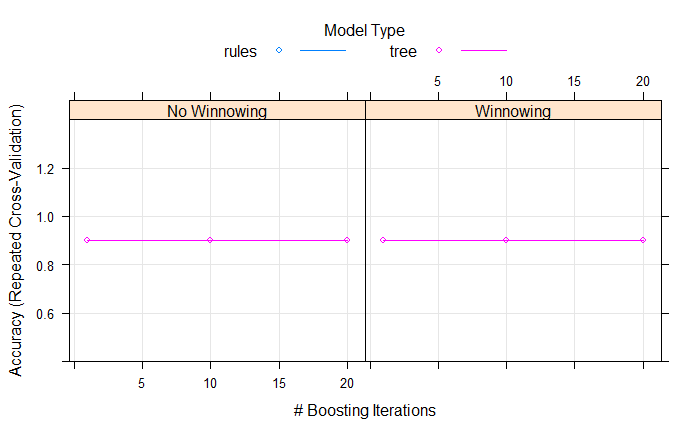}
\caption{C5.0}
\end{subfigure}
\begin{subfigure}[b]{0.45\textwidth}
\includegraphics[width=\textwidth]{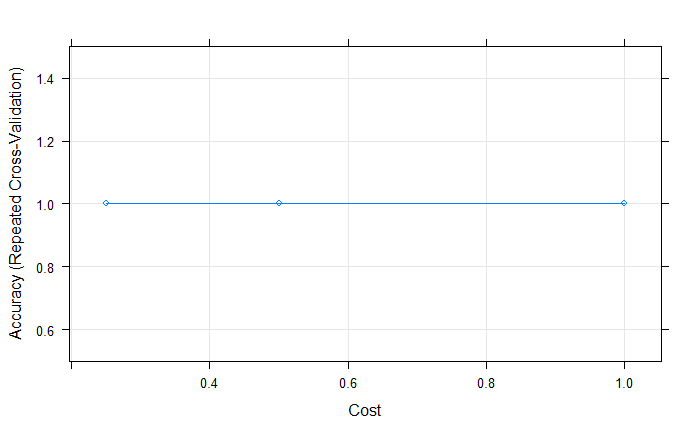}
\caption{SVMLinear}
\end{subfigure}
\begin{subfigure}[b]{0.45\textwidth}
\includegraphics[width=\textwidth]{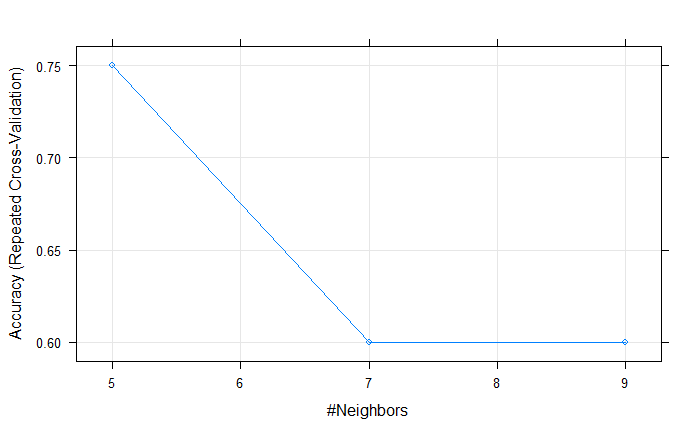}
\caption{KNN}
\end{subfigure} 

\caption{Performance and tuning results of training machine learning models on 80\% (10-fold cross-validation) of the Email data }
\label{fig:rq1_train}
\end{figure*}

We used the ``link prediction'' package in R \cite{lproximity} to calculate the eight similarity metrics on the Email product line's feature interaction graph.  Fig. \ref{fig:framework} shows how the feature interaction graph serves as input for the calculation of similarity indexes for each pair of nodes in it. Each of the eight similarity scores for a node is a variable in the data frame subsequently used to build the machine learning models. 

We next built a fully connected version of the graph.  Since there are 7 features in the feature interaction graph in Fig. 3, there are $(42/2)=21$ edges in the fully connected graph: the 10 in Fig. \label{fig:emailinteractiongraph} with the unwanted feature interaction label and the other 11 with the wanted feature interaction label. We use the labels for these 21 edges to train a supervised learning model to {\it classify an edge as an unwanted or wanted feature interaction}.

To do this, we selected six widely used machine learning algorithms to build models on the existing data, with the goal of  detecting missing or new unwanted feature interactions in a new product or new version of a system. These six ML algorithms were Neural Net (NNET), Naive Bayes (NB), Linear Support Vector Machine (SVMLinear), Decision Tree (C5.0), Random Forest (RF), and K- Nearest Neighbour(KNN). We used the following settings in our experiments to train, tune, and identify feature importance, and to test the data:
\begin{itemize}
    \item ``link prediction'' package in R to calculate the similarity metrics \cite{lproximity}.
    \item `` Caret'' package in R to train, tune, and test the machine learning models \cite{kuhn2020package}.
    \item 10-fold cross validation to train the models
    \item stratified splitting to divide the data into the 80\% training set and the 20\% test set. Stratified sampling ensured that the training and test sets have approximately the same percentage of samples of each target class as the complete set.
    \item variable importance for Random Forest, Neural Net, and C5.0 in ``Caret'' to identify the most important ML features.
    \item ROC curve variable importance to identify the most important features for those models such as Naive Bayes, SVMLinear, and KNN which do not have built-in variable-importance functions.
    
\end{itemize}

\vspace{5mm}

\textit {\textbf{RQ1:} How effectively does link prediction help detect unwanted feature interactions in a software product line?}

 \textbf{\textit{Training.}} Fig. \ref{fig:rq1_train} shows the performance results of 10-fold cross validation on the training data for the six machine learning algorithms. We report the Accuracy of the final tuned model for each machine learning algorithm. Fig. \ref{fig:rq1_train} also shows the tuning parameter and selected parameters for each machine learning model.  As shown there, the Naive Bayes model has an accuracy of 1 when we use Gaussian for the distribution type. Random Forest has the highest accuracy of 1 even using two randomly selected predictors. Neural Net has at most an accuracy of 0.95 when we use 1 hidden layer and weight decay of $1e-04$. KNN has at most an accuracy of 0.75 when we use $k=5$. Boosting Decision Tree (C5.0) has at most an accuracy of 0.90. SVM Linear has the highest accuracy of 1 with a mis-classification cost of 0.25.   
 
 \textbf{\textit{Testing.}} The results of the final models for each of the six machine learning models on 20\% of unseen data (i.e., two edges with unwanted feature interactions and two edges with wanted feature interactions) are shown in Table \ref{tab:test-result}.  We see that KNN mis-classified one of the two unwanted feature interaction as a wanted feature interaction. However, the other five machine learning algorithms  correctly detected all unwanted feature interactions on unseen data.
 
  \begin{table}[h]
  \caption{Performance Results of final trained machine learning models on 20\% of unseen Email data}
  \centering
    \begin{tabular}{ccccc}
    \hline
     & Model Name & Accuracy & Sensitivity & Specificity \\
    \hline
    & Naive Bayes  &1 &1&1\\
    & Random Forest &1 &1&1\\
    & Neural Net   &1 &1&1\\
    & Decision Tree (C 5.0) & 1  & 1&1 \\
    & SVM Linear &1 &1&1\\
    & KNN & 0.75  &0.5&1\\
    \hline
  \end{tabular}
  \label{tab:test-result}
\end{table}
 
The results indicate that using the machine learning algorithms along with link prediction helped detect new or missing unwanted feature interactions efficiently in the early development of a new product in a software product line.

\textit {\textbf{RQ2:} Which similarity metrics and machine learning algorithms perform better in the context of unwanted feature interaction detection?}

With regard to ML algorithms, the results described above showed that Random Forest, Naive Bayes, and SVM Linear had the highest Accuracy of 1 in classifying the feature interactions in the Email product line. These three machine learning models are all interpretable, simple and efficient.  

It is an open question at this point whether any of the other three models will be useful when we evaluate our approach on much larger datasets. For example, Neural Net generally performs better on large datasets so may be relevant for more complicated software product lines. 

\begin{figure*}[h]
\centering
\begin{subfigure}[b]{0.47\textwidth}
\includegraphics[width=\textwidth]{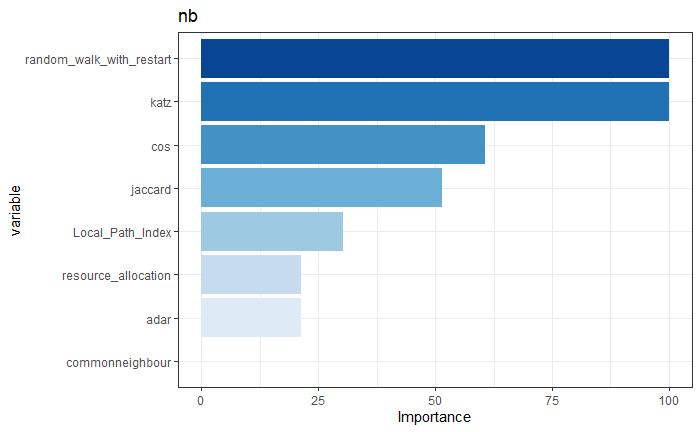}
\caption{Naive Bayes}
\end{subfigure}
\begin{subfigure}[b]{0.47\textwidth}
\includegraphics[width=\textwidth]{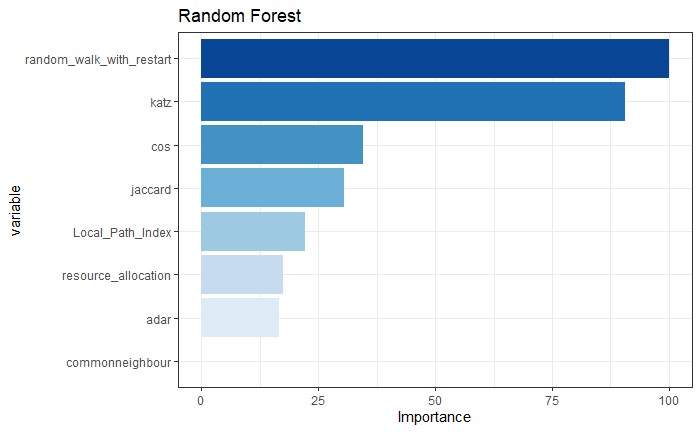}
\caption{Random Forest}
\end{subfigure}
\begin{subfigure}[b]{0.47\textwidth}
\includegraphics[width=\textwidth]{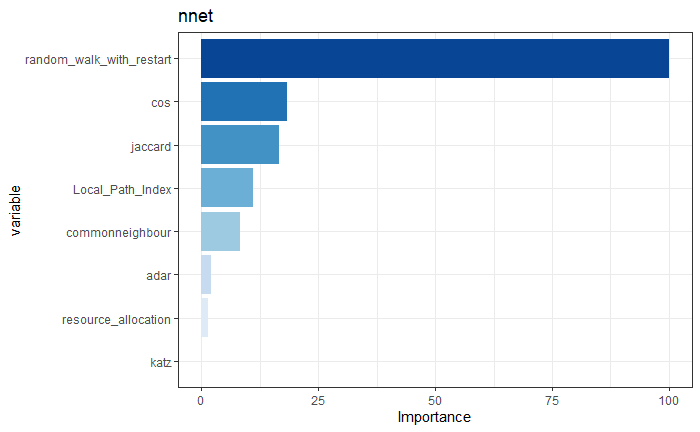}
\caption{Neural Net}
\end{subfigure}
\begin{subfigure}[b]{0.47\textwidth}
\includegraphics[width=\textwidth]{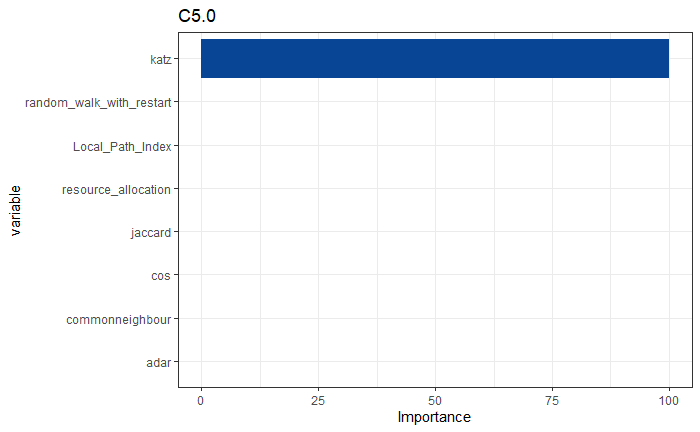}
\caption{C5.0}
\end{subfigure}
\begin{subfigure}[b]{0.47\textwidth}
\includegraphics[width=\textwidth]{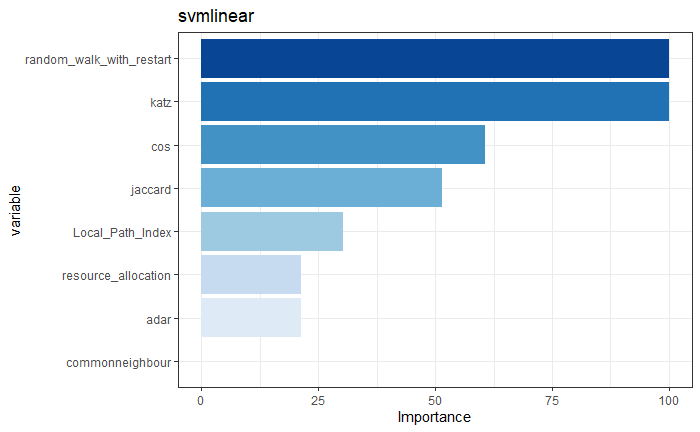}
\caption{SVMLinear}
\end{subfigure}
\begin{subfigure}[b]{0.47\textwidth}
\includegraphics[width=\textwidth]{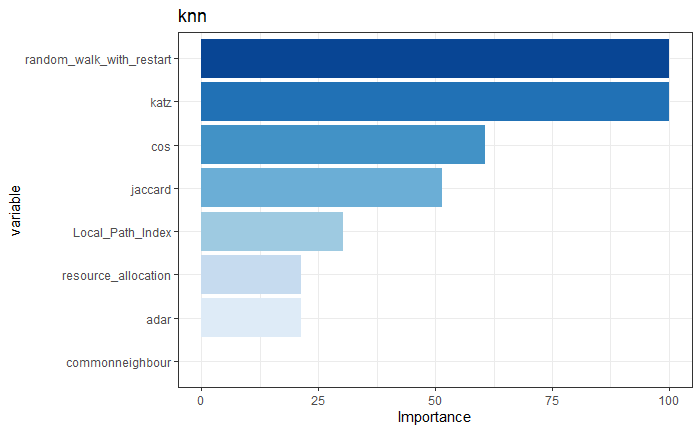}
\caption{KNN}
\end{subfigure}
\caption{Variable Importance of ML models in detecting unwanted feature interactions in Email}
\label{fig:rq2_email}
\end{figure*}

To answer which similarity scores perform better, we first extracted the feature importance related to each machine learning model described in RQ1. Fig. \ref{fig:rq2_email} shows the feature importance plot for these six machine learning models.  We used AUC (Area Under the ROC Curve) to extract the feature importance of Naive Bayes, KNN, and SVMLinear. We used the built-in variable importance function for Random Forest, Neural Net, and C5.0. 

The most important feature for all  models except C5.0 was ``Random walk with restart,'' which is a global similarity metric. ``Katz'' was the most important variable for C5.0 and was the second most important variable for Random Forest, SVM Linear, KNN and Naive Bayes. ``Katz'' is also a global similarity metric.  The next most important variables were ``Cosine'' and ``Jaccard'' which are local similarity metrics. 

The global similarity metrics thus played a more important role than local similarity scores in the link prediction.  Within global similarity metrics, ``Random walk with restart'' and ``Katz'' were the most important features. Within local similarity metrics,  ``Cosine'' and ``Jaccard'' performed better compared their peers.

\begin{figure}[h]
  \centering
  \includegraphics[width=\linewidth]{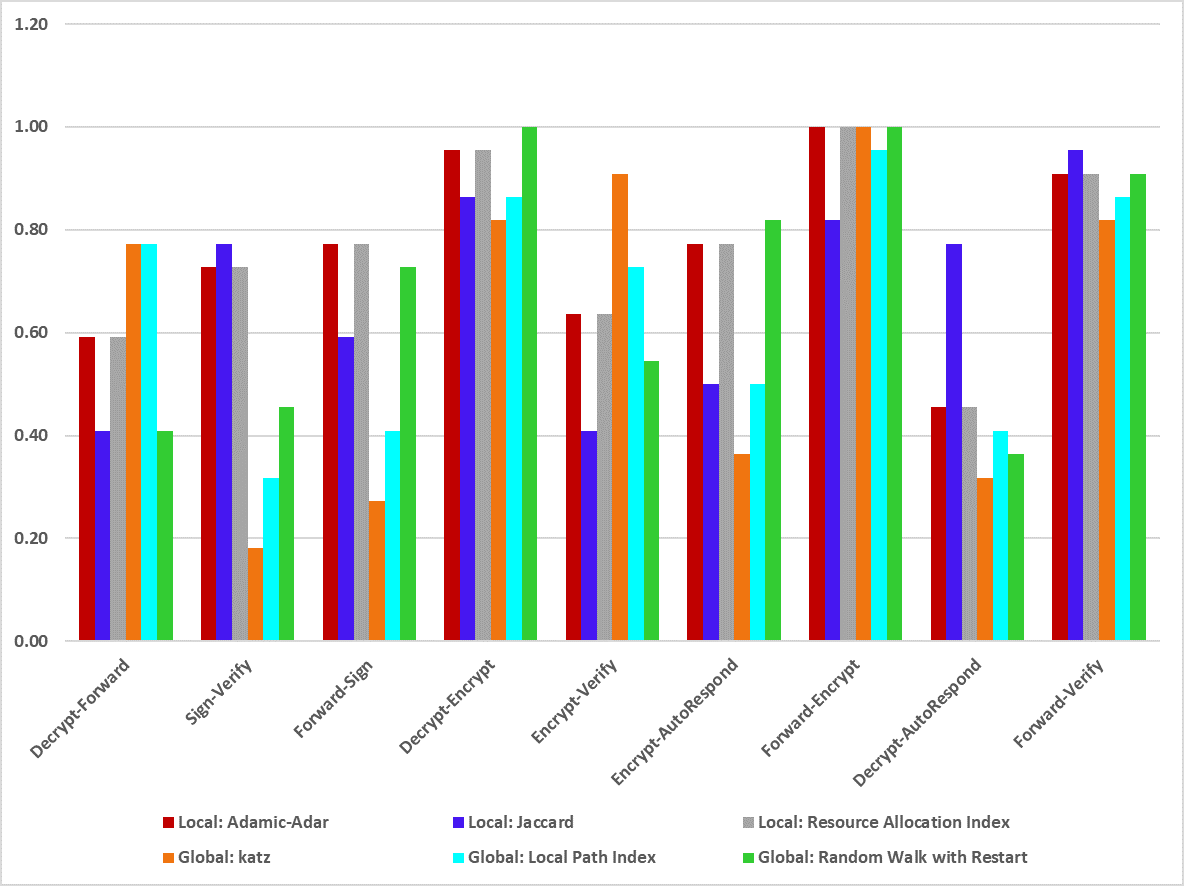}
  \caption{AUC (Area under the ROC Curve) of 6 different similarity metrics in detecting new unwanted feature interactions}
  \label{fig:detection-count}
 \end{figure}

Fig. \ref{fig:detection-count} shows the AUC (Area under the ROC Curve) bar chart of the six similarity scores in detecting each unwanted feature interaction pair in the Email system. AUC provides an aggregate measure of performance across all possible classification thresholds. As an example, to detect the unwanted feature interaction between ``Forward-Verify'', we calculated the similarity score of 6 different similarity metrics on the 9 remaining unwanted feature interactions and predicted the new edge ``Forward-Verify''.  The highest similarity score could detect the formation of the new pair.  For ``Forward-Verify'', the  ``Jaccard'' metric had the highest Area Under the ROC Curve, so performed better compared to other similarity metrics in detecting ``Forward-Verify''.  We see from  Fig. \ref{fig:detection-count}  that each of ``Katz,'' ``Random Wail With Restart,'' and ``Jaccard'' could detect 3 unwanted feature interactions, thus were more important compared to other similarity metrics for detecting unwanted feature interactions in the Email product line.

\textbf{\textit{Threats to validity:}}

{\it Internal threats.}  We have investigated a single, small case study as an initial investigation into the feasibility of our approach. However, this case study is considered to be a benchmark in the SPL literature and to have correct artifacts based on realistic wanted and unwanted feature interactions, so results on it are a good first step. 

The work described in this paper aims to detect unwanted pairwise feature interactions, where the presence of one of two features causes a change in the behavior of the other feature \cite{batory2011feature}.  This raises the question of whether our approach misses many interactions, namely those involving more than two features.  However, studies have shown that two-feature interactions are the most common form of feature interactions in software product lines  \cite{williams2000determination,siegmund2012predicting,oster2010automated}. A study analysing variability in 40 large software product lines also found that structural interactions exist mostly between two features \cite{liebig2010analysis}. Therefore, while more study is needed to generalize the results, our method's approach to handling pairwise feature interactions targets most, if not all, feature interactions \cite{batory2005feature,batory2011feature}.

{\it External threats.} While our work can be applied to other domains generally, we do not have any information about the applicability of this approach to large, real-world software product lines, and more work is necessary to confirm and improve these initial results.   Evaluation on additional software product lines in other application areas beyond that presented here also is needed.  Future work will seek to ascertain whether our proposed method for link prediction-based learning models for unwanted feature detection holds up and can be usefully applied to software product lines in other domains.

\section{Related Work}

Regarding the use of features as requirement engineering artifacts, Classen et al. \cite{classen2008s} defined a ``feature'' as a 
problem-level feature, which includes a set of related requirements, specifications and domain assumptions. They proposed a verification tool for Software Product Lines to uncover feature interactions. Our method differs from their framework in not using formal methods and model verification tools. Instead, we use machine learning algorithms to efficiently uncover unwanted feature interactions early in the development of a proposed new product in a software product line or of a new version in a software-intensive system.

In the area of detecting unwanted feature interaction using formal methods, Apel et al. \cite{apel2013feature,apel2011detection} introduced the ``feature-aware verification'' method to detect feature interactions using variability encoding automatically. Our approach to dealing with the feature interaction problem differs from their studies in exploiting known unwanted feature interactions and not requiring formal methods.

Atlee, Fahrenberg, and Legay\cite{atlee2015measuring} proposed an approach to measure the degree to which features interact in a software product line. They used transition systems and similarity metrics to compute the degree of feature interaction in a featured transition system. Our study differs from theirs in not requiring developers to produce a formal model of the system.

In the area of using similarity measures in a software product line, Henard, et al. \cite{henard2014bypassing} used similarity measures to prioritize test cases to decrease the number of product configurations in software product-line testing. Our study is different in that we used link prediction to detect unwanted feature interaction during requirements analysis, prior to coding or testing. 

Al-Hajjaji et al. \cite{al2014similarity} suggested a similarity-based prioritization that enhances coverage of SPL test cases to detect errors in a reasonable time. They compared the result of their algorithm with three sampling algorithms and concluded that the similarity-based prioritization algorithm could compete with them and produce the test cases faster.

S{\'a}nchez, Segura, and Ruiz-Cort{\'e}s \cite{sanchez2014comparison} investigated five different prioritization criteria including dissimilarity to generate test cases for software product-line testing. They obtained 87\% accuracy with prioritization based on dissimilarity. While we use a similarity model to detect feature interaction, their work differs from ours in that we apply similarity to individual features rather than to the entire product in a software product line and detect feature interaction during the requirements phase rather than the testing phase.

In the area of using link prediction, Lu and T.Zhou \cite{lu2011link,zhou2009predicting} examined the usage of local similarity measures based on the node similarity in link prediction on six real networks. They described the usage of link prediction methods in finding the missing links in the network and classification of partially labeled networks.

Rawashdeh and Ralescu \cite{rawashdeh2015similarity} investigated the structural and semantic similarity metrics in social networks. They compared different similarity metrics based on time and space complexity and highlighted the difficulty of choosing an appropriate similarity metric for link prediction. We are not aware of previous studies of link prediction for detecting missing or new feature interactions in a software product line.

We instead aim by our approach to use existing requirement artifacts, including knowledge of prior unwanted feature interactions and the software product line's feature model. We calculate similarity indexes and apply link predictions, then feed the results to machine learning algorithms to uncover unwanted feature interactions efficiently at an early stage of development of a new product in an evolving product line or new version in an evolving system. An open question is whether this method also could help developers converting a legacy system to a software product line identify unforeseen, problematic feature interactions.

\section{Conclusion}
This paper described a framework which uses machine learning algorithms along with a novel link-prediction based method to detect potential, unwanted feature interactions during the requirements phase of a new product or version in a software product line.  
 
Representing the software product line as a feature interaction graph enabled us to use  similarity-based link prediction and machine learning algorithms to detect unwanted feature interactions  much earlier in the development process for a new product than current testing techniques.  Results from application and evaluation on a small product line  showed that the best ML algorithms achieved good accuracy (0.75 to 1) for classifying product-line feature interactions as unwanted or wanted. 
Directions for future work are to evaluate our link-prediction method on a larger product line and to investigate whether incorporating structural similarity measures, as in \cite{khoshmanesh2018role}, improves classification of feature interactions.

\section*{Acknowledgment}
The first author thanks professor James Bailey for a useful discussion about link prediction. The work in this paper was partially funded by the US
National Science Foundation Grant CCF:1513717.

\bibliographystyle{plain}
\bibliography{ms}

\end{document}